# PHOTOSENSITIVITY OF $La_2CuO_4$ IN THE VICINITY OF THE PHASE BOUNDARY ANTIFERROMAGNET-SPIN GLASS.

Alexander A. Milner[*][†]

*Department of Physics & Astronomy, University of British Columbia, Vancouver BC, Canada*

The spectral, temperature and magnetic field dependencies of the linear dichroism in the reflection of light from the surface of an undoped single crystal of $La_2CuO_4$ are studied. The changes of the dichroism absolute value and its response to the external magnetic field induced by a weak optical illumination were detected. The set of experimental results is analyzed from the point of view of the high sensitivity of optical and magnetic properties of cuprates to the density of charge carriers localized in $CuO_2$ planes. It is assumed that the poor reproducibility of some of the observed effects is related to the proximity (on the x-T diagram of magnetic states) to the antiferromagnet – spin glass phase boundary.

Continuous attempts to explain high-temperature superconductivity and to increase $T_C$ stimulate interest in the state diagrams of the initial cuprates. In developing theoretical models involving the magnetic mechanism of pairing of charge carriers [1], attention is focused on a curious feature of the x-T phase diagram — the so-called transition from the antiferromagnetic (AFM) to the spin-glass (SG) state upon cooling of weakly doped non-conducting compounds based on $La_2CuO_4$ [2,3]. Briefly, the reason for this transition is as follows. It is known that AFM ordering of copper spins takes place due to the indirect exchange through $O^{2-}$ ions. Freezing a hole on oxygen changes the sign of the exchange interaction of the latter with copper, which causes frustration of the AFM bond at the site. At a certain concentration of holes, the long-range magnetic order is destroyed and the transition to the SG phase occurs. In this case, mutual attraction of holes, or rather, spatial formations of a hole–disturbed magnetic order is possible [2]. While there are numerous theoretical studies of the properties of quasi-two-dimensional frustrated antiferromagnets [4–7], convincing evidence has also accumulated in the experimental literature in favour of the AFM–SG phase transition in various systems [8–10].

Among the experimental methods for studying phases and transformations in x-T coordinates, photoexcitation methods are especially powerful. Their appeal is explained, in particular, by the maximum efficiency of varying the concentration of carriers. As applied to HTSC, the most illustrative in this regard is the photoinduced transition to the metallic state and photo-activation of superconductivity in YBa$_2$Cu$_3$O$_{6.4}$ films [11], as well as a noticeable increase of T$_C$ in laser-irradiated films [12] with unusually long lifetime of photoinduced state.

---

[*] amilner@phas.ubc.ca
[†] This work has been done in 1993 at the B. Verkin Institute for Low Temperature Physics and Engineering of the Ukrainian Academy of Sciences, Kharkov, Ukraine. However, it has never been published in its entirety.

This work is devoted to optical studies of the surface of a La$_2$CuO$_4$ crystal with a low hole concentration. I believe that by repetitively exposing the sample to optical radiation and heating, I was able to realize and record reversible changes in this concentration, apparently accompanied by changes in the magnetic subsystem of the crystal. Poor reproducibility of the observed effects can be related to the fact that events on the magnetic state diagram occur in the vicinity of the antiferromagnet–spin glass phase boundary. The state of the surface layer of the sample was monitored by the reflectance linear dichroism (RLD). Preliminary results of this study were reported previously in [13].

**Experimental technique and sample.**

The experiments were performed in an optical setup based on a polarization modulation technique for measuring the intensity ratio of the reflected light of two linear orthogonal polarizations, R1/R2. The block diagram is depicted in Fig.1.

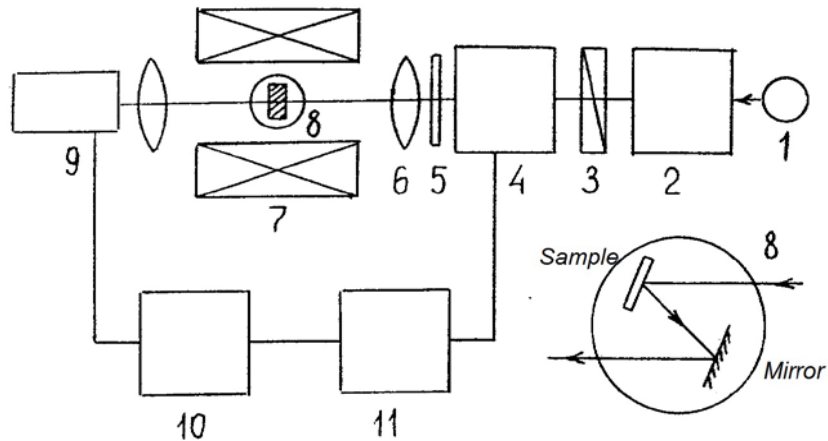

Fig.1. Experimental setup. 1 – source of white unpolarized light; 2 – monochromator; 3 – polarizer; 4 – piezo-optical modulator of polarization; 5 – neutral-density filter; 6 – lens; 7 – superconducting magnet; 8 – sample; 9 – photo detector; 10 – lock-in amplifier; 11 – controller of piezo-optical modulator.

The sample was mounted on a cold finger in the vacuum volume of a helium cryostat inside a superconducting magnet. The adjustable parameters were as follows: temperature 10 - 300 K, magnetic field 0 - 7 T, irradiation light wavelength 300 - 800 nm (1.5 - 4 eV). The light polarization modulation frequency by a piezoelectric quartz modulator was 36 kHz. Light intensity was varied from $10^{-5}$ to $10^{-7}$ W/cm$^2$ using neutral density filters (these values were determined with an accuracy of one order of magnitude at a wavelength of 625 nm). Almost all (except the one shown in Fig. 5) RLD measurements were performed at a minimum intensity of $10^{-7}$ W/cm$^2$. The illumination time at a maximum brightness, hereafter referred to as "exposure", typically took 20 minutes. The experiments normally lasted several days, beginning with a slow cooling of the sample from room temperature down to 10 K.

In this work, I studied the reflection of light from a mechanically polished surface, 1×1 mm², of one single crystal $La_2CuO_4$ with crystal lattice parameters $c$=13.146 Å, $a=b$=3.811 Å (Hereafter, the system of axes of the tetragonal phase is used). In accordance with the X-ray diffraction analysis, the angles between $c$, $a$ and $b$ axes and the reflective surface of the sample were 19, 15 and 65 degrees, respectively. RLD was measured for two mutually perpendicular polarizations, approximately coinciding with the directions of the projections of $a$ and $c$ axes on the surface, i.e. R1/R2 ≈ R$a$/R$c$. The angle of incidence of light on the sample was 14±1 degree (see Fig. 1). To take into account the effect of oblique incidence on RLD measurements, the experiments were carried out for two sample orientations, different by a rotation of π/2 around the normal to the surface: the plane of incidence contained either the crystallographic direction $c$ or $a$.

**Experimental results**

The RLD spectra at low and room temperatures are depicted in Fig. 2.

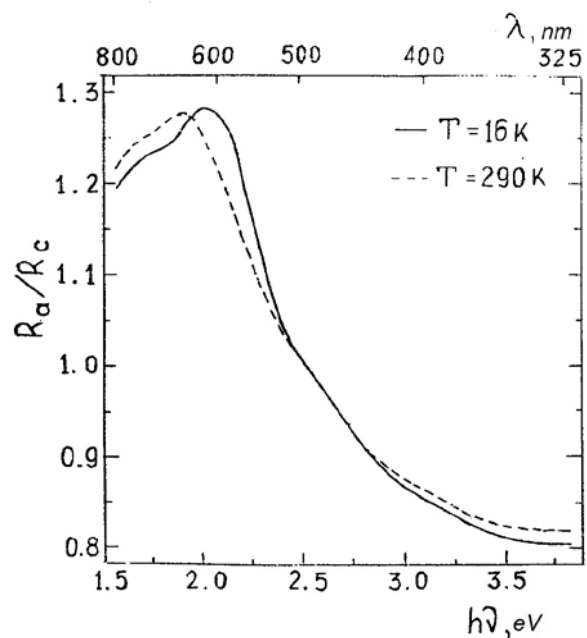

Fig.2. Spectrum of the reflectance linear dichroism (RLD) of $La_2CuO_4$.

At T= 6 K, the maximum appears at 2 eV. The temperature dependencies of R$a$/R$c$ at different frequencies are shown in Fig. 3. Their comparison with the curves in Fig. 2 indicates that upon cooling from room temperature to ~ 20 K, the main feature in the behaviour of the low-frequency part of the spectrum is the smooth shift of the dichroism band towards high energies by 0.12 eV. In the vicinity of 250 K, weak features are visible in the temperature dependencies, which cannot be explained by a change in the rate of temperature shift of the entire band or by the structure of its profile. Most likely, upon cooling in the region of 260 – 230 K, the band with a maximum near 2 eV weakens slightly, i.e. the shift in this temperature range is accompanied by a small decrease in amplitude, approximately by 0.2 – 0.3%.

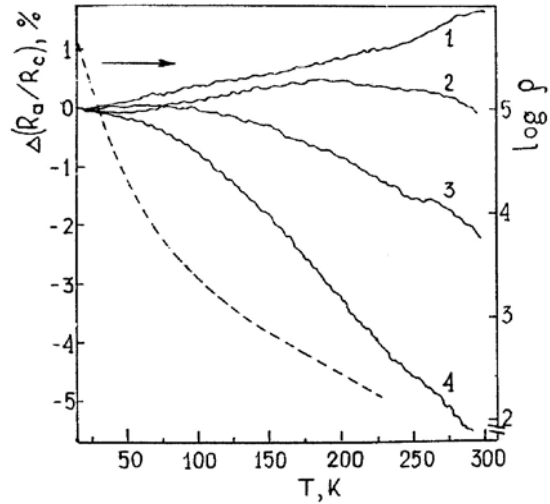

Fig.3. Temperature dependencies of RLD at different frequencies: 1 – 1.77 eV; 2 – 1.91 eV; 3 – 2.0 eV; 4 – 2.16 eV. Dashed line shows electrical conductivity of the sample.

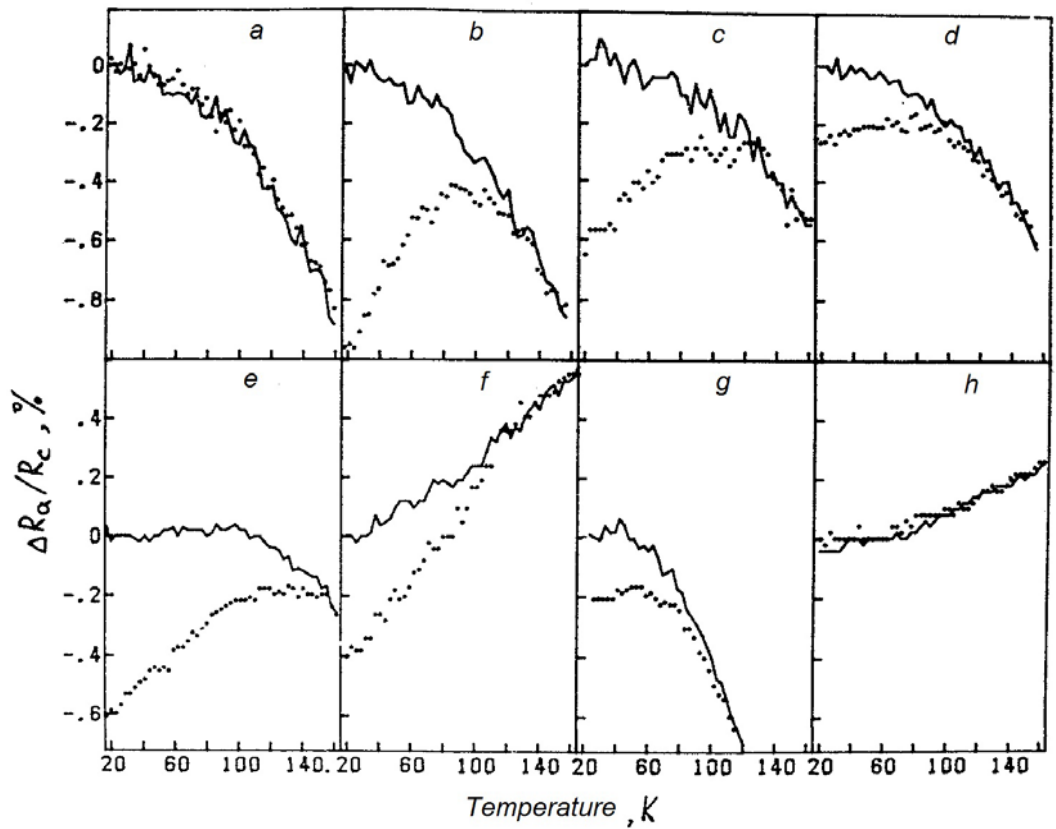

Fig.4. Temperature dependencies of RLD at different frequencies: a-e – 2 eV (repetitive measurements, see text for details); f – 1.91 eV; g – 2.14 eV; h – 3.54 eV. a – without pre-exposure, b-h – with pre-exposure. Solid lines – cooling, dots – heating.

Another feature of the $R_a/R_c$(T) dependencies is shown in Fig. 4. At temperatures up to 150 K, their appearance in the range from 1.5 to 2.3 eV was determined by the experimental sequence. If, after the previous heating-cooling, the sample was not exposed to light, the RLD monotonically and reversibly changed with temperature, as shown in Fig. 4a. If on the other hand it was exposed at T≤15 K, the course of the temperature dependencies during heating changed significantly, Figs. 4b – 4g. The exposure reduced the anisotropy by several tenths of a percent (Fig. 5).

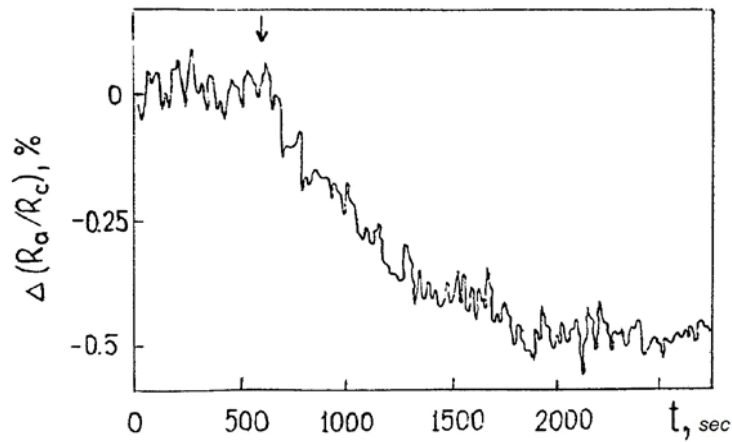

Fig.5. Time dependence of RLD at frequency 2 eV and T=15K. The arrow shows the moment of switching the light intensity from $10^{-7}$ to $10^{-5}$ W/cm$^2$.

The cycles of cooling-exposure at T≈15 K for ~ 20 minutes-heating to 140–150 K, could be repeated many times, while the qualitative behaviour remained similar. However, it is worth emphasizing that the quantitative details of the experimental curves $R_a/R_c$(T) were not reproducible at T≤150 K. Namely, the growth of RLD upon heating of the exposed sample could begin both immediately at a minimum temperature of 15 K and with a greater or smaller temperature delay. Also, at the frequency of 2 eV, it reached maximum at a temperature different from measurement to measurement, 80 ÷ 100 K, and when cooled from 150 K, it most often increased smoothly, although sometimes it reached its highest value in a rather narrow temperature range. The difference in RLD at T = 15 K at the beginning and the end of the "exposure–heating–cooling" cycle was not the same, too. Some characteristic dependencies are presented in Fig. 4b – 4d. I emphasize that the indicated irreproducibility could be observed within a single series of measurements, even in two temperature cycles following one another. It was found that a decrease of RLD could be initiated by light of any frequency in the accessible range of 1.5 - 4 eV, regardless of its polarization, i.e. either unpolarized or with $\mathbf{E} \parallel c$ and $\mathbf{E} \perp c$. The influence of the exposure to light was not manifested in the short-wavelength part of the RLD spectrum, at hν ≥2.3 eV (Fig. 4h). A comparison of curves in Fig. 4, related to different frequencies, with each other and with the RLD spectrum in Fig. 2, shows that a decrease in the reflection anisotropy under the influence of light is associated with a decrease in the amplitude of the band with a maximum of 2 eV, and not with its frequency shift.

Each of the surface states, before and after exposure, turned out to be very long-lived. Temperature dependencies taken 16 hours after the previous experimental sequences, whether without or with pre-exposure, had the character, respectively, as in Fig. 4a or 4b.

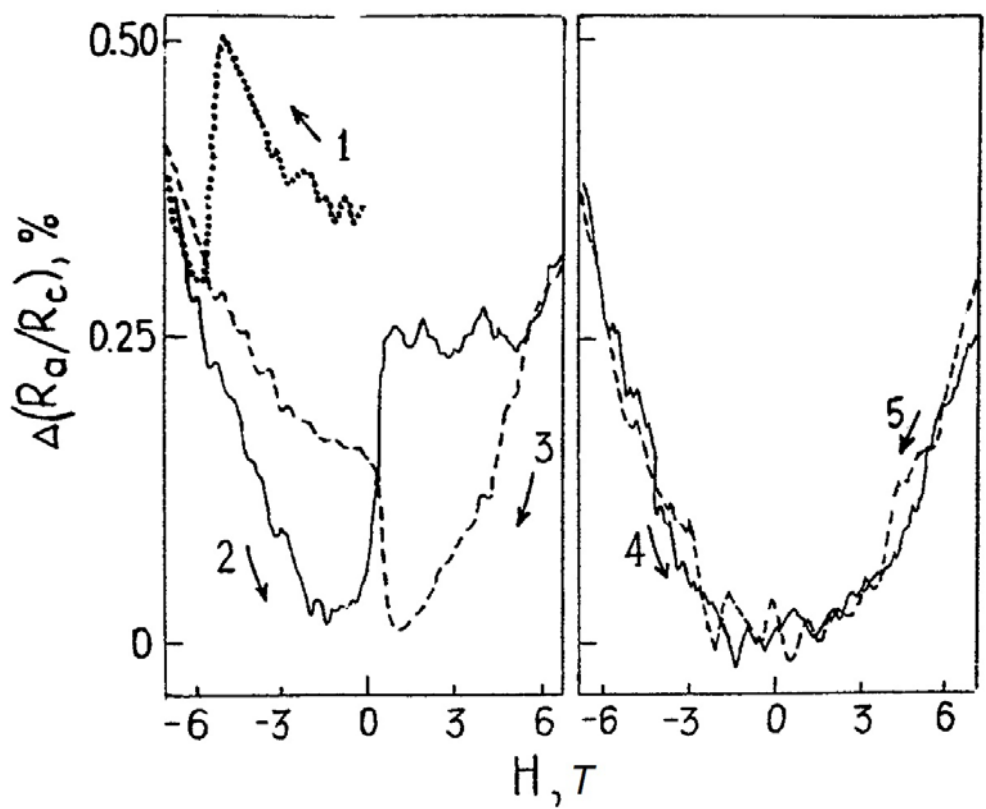

Fig.6. Magnetic field dependencies of RLD of pre-exposed $La_2CuO_4$ at the frequency of 2 eV. Numbers and arrows depict the sequence of measurements, separated for clarity in two plots.

In addition to temperature, the magnetic field dependencies of RLD were also sensitive to the action of light, although in this case the reproducibility of the effect was much worse. The effect consisted in the appearance of jumps on the $Ra/Rc(H)$ curves after the exposure to light in one of the measurement series. RLD was recorded during the slow scanning of the magnetic field in the range of ± 7 T. Figure 6 shows the sequence of curves obtained in low-intensity light at T = 15 K, hν=2 eV, after exposing the crystal to bright light. The next two exposures again led to the appearance of jumps in the $Ra/Rc(H)$ dependence, which, as in the first case, disappeared after 2–3-fold cycling of the magnetic field. Heating in the described sequence of experiments was not performed. Then the sample was heated to 225 K, cooled and aged at 15 K for several hours. Although after this temperature cycle the crystal was not exposed to bright light, the magnetic field dependences turned out to be similar to those shown in Fig. 6, i.e. contained jumps that disappeared after several field scans. In subsequent measurements, the above effect was observed twice again; in further experiments of this and other series, neither intense light nor temperature variations led to the appearance of jumps in the $Ra/Rc(T)$ curves. It is worth noting that when the sample was rotated by 90 ° (as mentioned above, this was done to take into account the oblique incidence of light), the dependences looked different, due to the changed projections of the magnetic field on the crystal axes.

**Discussion**

In the frequency range from 1.5 to 5 eV, the optical spectrum of $La_2CuO_4$ is dictated by the transitions between bands formed from 2p oxygen states and 3d copper states [14, 15]. The maximum of RLD near 2 eV is apparently related to the peculiarity in the reflection spectrum observed in the $E \perp c$ polarization and attributed to the transitions with charge transfer from O(2p) to Cu(3d) in $CuO_2$ planes [14, 16, 17]. It was found that the amplitude of the reflection peak at the frequency of 2 eV at $\mathbf{E} \perp c$ is very sensitive to the dopant concentration in compounds based on $La_2CuO_4$ and, accordingly, to the density of electric charge carriers (see, for example, Fig. 6 in [17]). Their electric field lowers the probability of these transitions [14, 17]. A pronounced feature with significant amplitude in the RLD spectrum (Fig.2) indicates a low carrier concentration. Therefore, since the supplier of holes in the $CuO_2$ plane can be the excess of oxygen [18], the oxygen index in the compound under study is really close to four (Caution in relation to the adequacy of the chemical formula is due to the well-known lability of this compound to oxygen, especially near surface). This conclusion is supported by the essentially semiconductor character of the temperature dependence of the electrical resistance of our sample (see Fig. 3) [19], as well as the peculiarity of the temperature dependence of RLD in the region of 250–260 K, if one agrees that it owes its origin to antiferromagnetic ordering. This is most likely the case, because, firstly, $T_N$ is characteristic for almost stoichiometric composition [19-22], and secondly, an anomaly of the thermal expansion coefficients of $La_2CuO_4$ is observed in the $T_N$ region [21]. The assumption of direct and indirect, through magnetoelastic coupling, contribution of the crystal magnetic subsystem to the anisotropy of interaction with light is quite reasonable. The smooth shift of the RLD spectrum toward higher energies during cooling (Fig.2) is apparently associated with a temperature change in the lattice parameters and reproduces the shift of the fundamental absorption edge typical for semiconductors [23].

Comparing the results presented in Fig.2 and 3 with known data [17-19,22], I conclude that the surface layer of the studied $La_2CuO_4$ single crystal was characterized by a relatively low density of free carriers $n_h \leq 5 \cdot 10^{19} cm^{-3}$ (which corresponded to the hole concentration on a $CuO_2$ complex of $p \leq 0.01$) and, accordingly, the AFM order at low temperature.

Turning to the discussion of the effect of photoinduced changes in optical anisotropy, Fig. 4, I highlight its main properties:

1) When a cold sample is exposed to light, RLD in the vicinity of the spectral maximum of 2 eV decreases by several tenths of a percent.

2) The effect is induced by light of any frequency in the range of 1.5 - 4 eV, regardless of its polarization.

3) Heating to $T \geq 100$ K restores the value of $Ra/Rc$.

4) The effect is accumulative. When exposed to light with an intensity of $10^{-5}$ W/cm$^2$, the characteristic saturation time is ~ $10^3$ sec. The life time of the photoinduced state is long, at least $10^5$ seconds.

5) The magnitude of the photoinduced decrease in RLD and its temperature dependence after the exposure to bright light are not reproduced exactly from measurement to measurement.

6) The exposure may uncontrollably alter the reaction of RLD to an external magnetic field.

Based on the significant physical equivalence of the role of charge carriers introduced in the CuO$_2$ plane by photoexcitation and chemical doping [24,25,26], I suggest the following nature of the described phenomenon.

Optical irradiation excites charge transfer transitions, both in the CuO$_2$ planes and between ions outside these planes. As a result, holes accumulate in the copper-oxygen planes. They create a screening electric field, which reduces the probability of light absorption with polarization $\mathbf{E}\perp c$. In this case, the reflection of R$a$ and, accordingly, the value of R$a$/R$c$ decreases.

Taking into account the strong dependence of R$a$ on the dopant concentration, for example, Sr (see [17]), and assuming that one doped strontium atom generates one extra hole in the CuO$_2$ layer [27], we see that to achieve the indicated attenuation of R$a$ one needs to place holes on less than 1% of planar oxygen atoms.

In a volume of $1\times1\times10^{-5}$ cm$^3$ ($10^{-5}$ cm is the penetration depth of light), there are ~ $10^{16}$ molecules. Given the light intensity of $10^{-5}$ W / cm$^2$, ~ $10^{13}$ photons interact with such a volume every second. Therefore, in the presence of a reliable mechanism of capture and long-term retention of holes, the saturation of the effect during $10^3$ sec can be explained with this mechanism.

In general, structural inhomogeneities, impurities (in particular, excess oxygen atoms), and other defects can serve as traps that provide carrier localization. This natural assumption, however, contradicts the fact of poor reproducibility of the effect. Perhaps in this case the decisive role is played by the processes of hole self-localization.

The basis for this assumption is the significant effect of electronic excitations on the crystal structure and the strong mutual influence of charge carriers and local magnetic moments that have been reliably established to date. The former of these effects underlies the interpretations of many photoinduced phenomena, including long-lived ones [24, 25, 26], the latter is involved in the analysis of magnetoresistance data [8, 28, 29] and in discussing the relationship between magnetism and superconductivity in high-temperature superconductors [1]. This work has insufficient data to answer the question about the mechanism of stabilization of the photoinduced effect. However, I consider the assumption that the magnetic subsystem of the crystal participates in the observed phenomenon well justified. In accordance with the above reasoning, as well as with data on the extremely high sensitivity of the magnetic order to hole concentration in the CuO$_2$ planes [19-21, 30, 31], our sample, being cooled and unexposed, is in a state close (on x-T diagram [32]) to the boundary between the AFM and SG phases. Holes appearing under the exposure to light, introducing local perturbations into the system of AFM-ordered copper spins, and possibly fixing themselves on these defective sites created by them, accumulate and gradually transfer the final surface volume of the crystal to the SG phase. Due to the possible interaction (attraction) between the perturbed centers, the process can be accompanied by the formation of inhomogeneous spatial structures (up to the effect on the crystal twin structure, as suggested in [21]; here one should bear in mind a very low anisotropy in the basal plane). This, in turn, can cause the irreproducibility of the effect. A strong argument in favor of such a magnetic scenario is the change in the magnetic field dependence of RLD under the influence of light. Results shown in Fig. 6 also indicate, possibly, a considerable magnetic contribution to the reflection anisotropy. This contribution should be sensitive to the orientation of magnetic moments and the degree of coherence of the spin system. Unfortunately, the complex geometry of the experiment, in the sense of the orientation of the magnetic field relative to the crystallographic directions, and mainly the uncontrolled and rarely occurring effects described above, did not allow me to answer the questions of how does the external magnetic field affect the frustrated AFM state, how do the ferromagnetic regions around the localized holes behave, whether the system becomes more uniform as a result of magnetic field cycling, etc.

Heating of the exposed sample leads to an increase in the mobility of carriers (it is known that the transport mechanism changes from hopping to diffusion in the region of 50–100 K [18, 20]) and the restoration of the long-range AFM order with simultaneous annihilation of holes and their diffusion deep into the crystal (recall that in our sample the $CuO_2$ plane, inside which the carrier mobility is the highest, is perpendicular to the surface under study). Therefore, taking into account the natural heterogeneity of the sample (variations of the oxygen index, density of defects, etc.), the entire described cycle can be represented on the phase diagram as shown in Fig. 7.

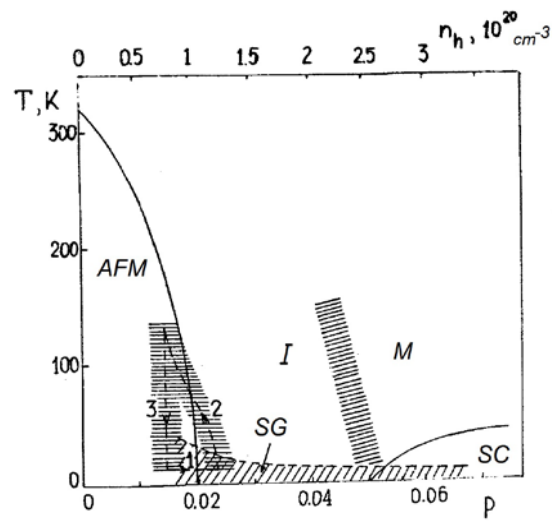

Fig.7. Probable trajectories in the cycle: exposure (1) – heating (2) – cooling (3) on the phase diagram, built on the basis of x-T diagram from [32]. At values x<0.2 the dopant concentration x in $La_{2-x}Sr_xCuO_4$ coincides with the number of holes P per complex $CuO_2$ [22]. Phases: AFM – antiferromagnet, SG – spin glass, SC – superconductor, I – insulator, M – metal.

Thus, the change in the surface state of the lanthanum cuprate under the influence of relatively weak light with intensity of the order of ~$10^{-5}$ W/cm$^2$, monitored in this work by the reflectance linear dichroism method, is caused by an increase in the number of holes localized at low temperature in copper-oxygen planes. At a low carrier concentration, this process can be accompanied by a violation of the long-range magnetic order and the formation of regions of the spin-glass phase. The interaction of the latter with one another, the interaction of the magnetic structure with the distorted crystal, the natural heterogeneity of the surface, and the likely magnetic contribution to optical anisotropy can all cause the observed irreproducibility in a number of experiments.

**References**

1. Emery V.I. - Phys.Rev.Lett., 1987, vol.58, N 26, p.2794.
2. Aharony A., Birgeneau R.J., Coniglio A. et al. - Phys.Rev.Lett., 1988, vol.60, N 13, p.1330.
3. Glazman L.I., Ioselevich A.S. - Z.Phys.B.Condensed Matter, 1990, vol.80, N 1, p.133.
4. Lozovik Yu.E., Notych O.I. – JETPH Lett., 1991, vol. 54, N 2, p.91.
5. Bogdan M.M., Kovalev A.S. – Fizika Nizkikh Temperatur, 1990, vol.16, N.112, p. 1576.
6. Loktev V.M. – Superconductivity: Physics, Chemistry, Technics (Moscow), 1991, vol.4, N 12, p.2293 (in Russian)
7. Paul P., Mattis D.C. - Phys.Rev.B, 1991, vol.44, N 5, p.2384
8. Zakharov A.A., Teplov A.A., Krasnoperov E.P. et al. - JETPH Lett., 1991, vol. 54, N 1, p.30.
9. Bazhan A.N., Bevz V.N. - Superconductivity: Physics, Chemistry, Technics (Moscow), 1991, vol.4, N 1, p.116 (in Russian)
10. Gezalyan A.D., Shul'pekova S.V. - JETPH Lett., 1991, vol. 54, N 1, p.47.
11. Kudinov V.I., Kirilyuk A.I.,Kreines N.M. et al. - Phys.Lett.A, 1990, vol.151, N 6,7, p.358.
12. Nieva G., Osquiguil E., Guimpel J. et al. - Appl. Phys. Lett., 1992, vol. 60, N 17, p.2159.
13. Milner A.A., Kharchenko N.F., Miroshnichenko V.A., Kosmyna M.B. – Superconductivity: Physics, Chemistry, Technics (Moscow), 1989, vol.2, N 7, p.98 (in Russian)
14. Tajima S., Ishii H., Nakahashi T. et al. - JOSA B, 1989, vol.6, N 3, p.475.
15. Varma C.M., Schmitt-Rink S., Abrahams E. - Sol.St.Commun.,1987, vol.62, N 10, p.681.
16. Eklund P.C., Rao A.M., et al. - JOSA B, vol.6, N 3, p.389.
17. Uchida S., Ido T., Takagi H, Arima T., Tokura Y., Tagima S.- Phys.Rev.B, 1991, vol.43, N 10A, p.7942

18. Preyer N.W., Birgeneau R.J., Chen C.Y. et al. - Phys.Rev.B, 1989, vol.39, N 16, p.11563.
19. Johnston D.C., Stokes J.P., Goshorn D.P. et al. - Phys.Rev.B, 1987, vol.36, N7, p.4007.
20. Cheong S.-W., Thompson J.D., Fisk Z. - Physica C, 1989, vol.158, N 1-2, p.109.
21. Cheong S.-W., Hundley M.F., Thompson J.D. et al. - Phys.Rev.B, 1989, vol.39, N 10, p.6567.
22. Chen C.Y., Birgeneau R.J., Kastner M.A. et al. - Phys.Rev.B,1991, vol.43, N 1, p.392.
23. Moss T.S., Burrell G.J., Ellis B. – Semiconductor Opto-Electronics, Buttrworth &Co. (Publishers) Ltd, 1973.
24. Kim Y.H., Foster C.M., Heeger A.J. et al. - Chem. of $HT_c$Supercond.11, 1988, ed. D.Nelson, p.194.
25. Ginder J.M., Roe M.G., Song Y. et al. - Phys.Rev.B, 1988, vol.37, N 13, p.7506.
26. Kim Y.H., Cheong S-W., Fisk Z. - Phys.Rev.Letters, 1991, vol.67, N 16, p.2227.
27. Torrance J.B., Tokura Y., Nazzal A.I. et al. - Phys.Rev.Lett., 1988, vol.61, N 9, p.1127.
28. Thio T., Chen C.Y., Freer B.S. et al. - Phys.Rev.B, 1990, vol.41, N 1, p.231.
29. Gogolin A.O., Ioselevich A.S. – JETP, 1990, vol.71, N2, p.380
30. Uchida S., Takagi H., Tokura Y. - Physica C, 1989, vol.162-164, p.1677.
31. Watanabe I., Kumagai K., Nakamura Y. et al. - J.Phys.Soc.Jpn., 1987, vol.56, N9, p.3028.
32. Birgeneau R.J., Shirane G, in "Physical Properties of High Temperature Superconductors", D. M. Ginsberg, Ed. (World Scientific, Singapore, 1989), vol. 1, p. 151.